# Evaluation of motion comfort using advanced active human body models and efficient simplified models

Raj Desai, Marko Cvetković, Georgios Papaioannou, Riender Happee

*Abstract*— Active muscles are crucial for maintaining postural stability when seated in a moving vehicle. Advanced active 3D non-linear full body models have been developed for impact and comfort simulation, including large numbers of individual muscle elements, and detailed non-linear models of the joint structures. While such models have an apparent potential to provide insight into postural stabilization, they are computationally demanding, making them less practical in particular for driving comfort where long time periods are to be studied. In vibrational comfort and in general biomechanical research, linearized models are effectively used. This paper evaluates the effectiveness of simplified 3D full-body human models to capture comfort provoked by whole-body vibrations. An efficient seated human body model is developed and validated using experimental data. We evaluate the required complexity in terms of joints and degrees of freedom for the spine, and explore how well linear spring-damper models can approximate reflexive postural stabilization. Results indicate that linear stiffness and damping models can well capture the human response. However, the results are improved by adding proportional integral derivative (PID) and head-in-space (HIS) controllers to maintain the defined initial body posture. The integrator is shown to be essential to prevent drift from the defined posture. The joint angular relative displacement is used as the input reference to each PID controller. With this model, a faster than real-time solution is obtained when used with a simple seat model. The paper also discusses the advantages and disadvantages of various models and provides insight into which models are more appropriate for motion comfort analysis. For designers and researchers in the automotive and seating industries, the findings given in this paper provide useful insights that will help them improve the comfort and safety of both vehicle occupants and seats.

## I. INTRODUCTION

Motion comfort [1], [2] is an important factor in designing vehicles, aircraft, and other transportation systems. It refers to the level of comfort that passengers experience during motion, including factors such as ride smoothness, motion sickness, whole-body vibration, and noise [3]. Evaluating motion comfort can be challenging, as it depends on a variety of factors, including vehicle design, road conditions, and human factors. One approach to evaluating motion comfort is to use advanced active human body models. These models simulate the behaviour of the human body during motion and can provide detailed information on postural stability, head acceleration a key factor in perceived comfort and motion sickness, muscular effort, forces and stresses that passengers experience. However, these models can be complex and computationally intensive, requiring significant computational resources to run.

By enabling users to perform other tasks while driving, automated driving (AD) has the potential to offer safe and environmentally friendly transportation. In contrast to conventional automobiles, this complicates the occupants' total postural stability even more [4]. Inconvenience and even low back pain or lumbar spine injury might result from the vehicle's whole-body vibration (WBV). As a result, human-centered design of automated driving systems [5], [6] requires knowledge of and models for human motion and perception. Therefore, it's crucial to comprehend how WBV affects the human body and how vibrations are transferred through it.

One approach to evaluate motion comfort is to use simplified, efficient models [7]–[9], which are much faster to run than complex human body models and can be useful for early-stage design evaluations or for evaluating a large number of design options. However, simplified models may not capture all details of the human body's response to motion. On the other hand, advanced active human body models like THUMS [10] and MADYMO [11] can provide highly detailed information on the forces and stresses that passengers experience but require specialized expertise and a large amount of computational time.

Experiments have limits in fully capturing the range of motion, even though they can provide useful information regarding human mobility [12]. Therefore, developing a computationally efficient and accurate 3D human body model would be a significant advancement in the field of human motion comfort analysis with important applications in various domains such as automotive design, aircraft cabin layout, public transportation systems, virtual reality experiences, and ergonomic product design. The model must also maintain the desired posture, which is another crucial criterion. This requires the use of active muscular feedback forces or torques. A PID (Proportional-Integral-Derivative) controller is widely used as a feedback control method in engineering and industrial applications [13]. The controller adjusts the control signal based on the difference between the desired setpoint and the measured process variable, which is also known as the error signal. The PID controller calculates the control signal by taking into account three parameters: proportional gain, integral gain, and derivative gain. The computationally efficient human model (EHM) presented in this paper employs a closed-loop PID feedback and HIS joint torque controllers for maintaining occupant's desired posture.

The EHM is built on the rigid body modelling capabilities of MADYMO. The inertial properties of the bodies are included in the rigid bodies of the model, and their geometry is described by ellipses and planes. Kinematic joints are utilized to organize the structural deformation of flexible components

*Research supported by Toyota Motor Corporation     Intelligent-vehicles group, Cognitive Robotics (CoR) TU Delft

under dynamic restraint models. The ellipsoids' force-based contact characteristics show how soft tissues, like skin and flesh, deform. These characteristics characterize the contact interactions between the human body and the seat. Among the MADYMO models, the MADYMO detailed active human model (AHM) represents the 50th percentile male population and has been validated for impact conditions [14], [15] and for vibration and dynamic driving [16]. The model geometry consists of standing height (1.76m), sitting height (0.92m) and weight (75.3kg) derived from the ergonomic model in Ramsis [5]. The controllers for the spine, neck, shoulders, elbows, hips, and knees make up the active human model. The skin is captured by finite element surfaces for contact interaction and there are 190 bodies in the AHM (182 rigid and 8 flexible). As a result, the AHM requires extensive computing time. We provide a computationally effective human model (EHM) for comfort analysis to lessen this for vehicle comfort simulation. The efficient MADYMO human body model will allow researchers to effectively explore human body responses to WBV without consuming large amount of computational time.

The EHM model is designed to be computationally efficient and simple, while accurately representing 3D body joint biomechanics and providing a good fit with experimental motion [12]. In building the model, a functional set of body segments is used, selecting only those that have a significant impact on the kinematics and dynamics of the body. Here we consider bodies head, trunk and pelvis, and examine 3D motion in translation (x-y-z) and rotation (roll-pitch-yaw). PID joint controllers are used to stabilize the posture. As a dependent function of posture, seat interactions must be represented in the model. The model, therefore, has accurate interactions with the floor, the seat base, and the back. To benchmark the EHM, the AHM is used for comparison of model performance. To our knowledge, there hasn't been a published 3D posture control multi-body human body model that has been thoroughly (6 DoF with head, pelvis, and trunk, vertical/fore-aft/lateral) validated. The purpose of this study is to develop a human body model that is efficient, flexible to capture subjective responses with changes in anthropometric data, maintain/achieve a desired posture, and can be applied to analyse human behaviour in an automated vehicle environment.

## II. BIOMECHANICAL MODELLING

In order to build an efficient seated human body model, reported biomechanical models in literature [7], [17] and MADYMO AHM were investigated. We evaluated the MADYMO AHM since we had the flexibility to adjust the model parameters to our experimental results. Fig. 1 [18] presents the AHM and the EHM in the configuration used for validation in this paper. This configuration is tested in [12] and in this paper we use the condition with erect posture with high support (lower support pad at posterior superior iliac spine, another pad aligned with the tip of the inferior scapula's angulus).

*Body segments and joints:* The pelvis, lower torso, middle torso, upper torso, neck, head, left thigh, right thigh, left lower leg, right lower leg, left foot, and right foot are the 12 segments that make up the EHM. The model is designed to predict in particular the trunk and head motion. In order to realistically predict how a seated human body responds to vehicle vibration, legs and feet are also added, as our previous research has shown relevant contributions of the legs in trunk stabilization in a dynamic slalom drive [16]. Therefore, the model must depict legs and how they interact with the ground.

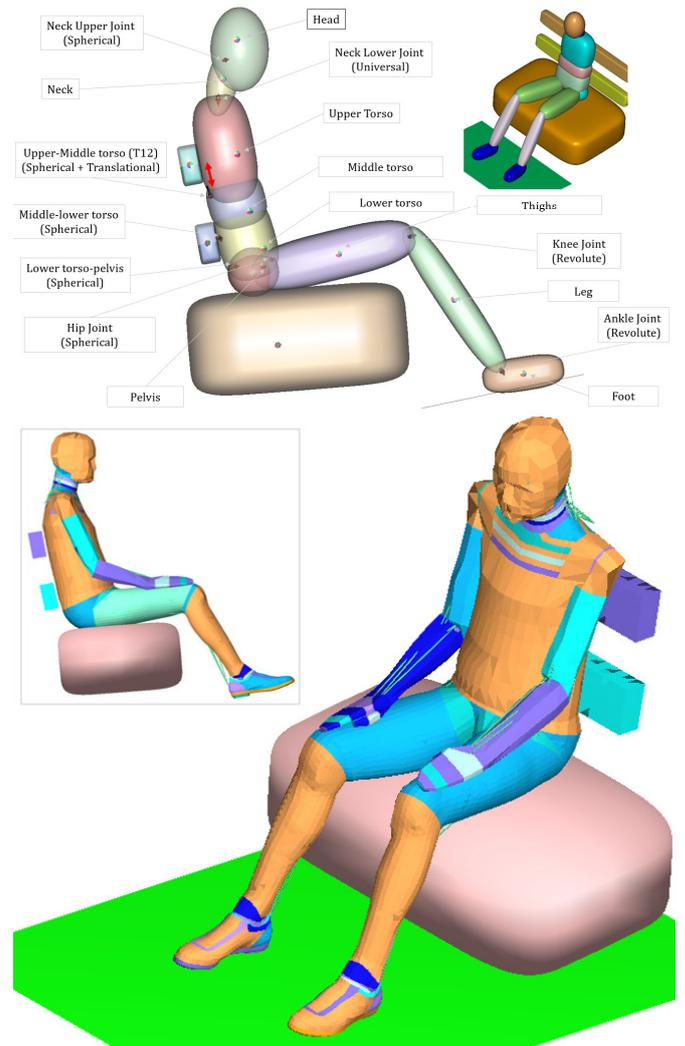

Fig. 1 EHM (upper) and AHM (geometry)

The various body parts are connected via kinematic joints. A spherical joint connects pelvis with lower torso while an additional spherical joint is placed between L4-L5 to capture lumbar bending [19] as this forms the rotational point between lower and middle torso. A spherical-translational joint connects the middle and upper torso and allows for free rotation and vertical movement. The spinal compression/extension during vertical loading is captured by this vertical joint degree of freedom. The spherical joints are utilized to simulate 3D rotation, which includes the torso's flexion-extension, abduction, adduction, and yaw rotation. A spherical joint is placed at upper neck located at (C1-C0) to capture the head yaw-pitch-roll and at the lower neck (T1-C7) a universal joint captures roll and pitch motion [20]. The right and left hip joints are also modeled as spherical joints to connect the thighs and pelvis, while the right and left knee joints are modelled as revolute joints, allowing for relative rotational movement around one axis between the thighs and lower legs. Ankle connections are modelled as revolute joints, and they connect lower legs and feet. By superimposing the two models, it can be shown that the EHM's head, trunk (T8), and pelvis all have centres of gravity (C.G.) that are situated close to those of the AHM. Fig. 1 depicts information concerning joints in detail. Taking into account the numerous rotational and translational movements permitted by the kinematic joints and their restrictions, the EHM model has 31 degrees of freedom (DoF).

*Seat interaction:* The model is validated using an experiment on an experimental compliant seat carried out by our research team [12]. Participants were instructed to sit in a mock-up car while motions were generated by random vibrations in the fore-aft, vertical, and lateral directions. The seat pan, backrest, and floor of the MADYMO model environment correspond to the three segments of the car mock-up. A plane serves as the floor, and ellipses represent seat pan and backrest. In order to facilitate comparisons, a male body size in the 50th percentile that is near to the mass of the average human model is used with size and inertia parameters from anthropometry measures found in the literature [21]. To capture the human-seat interaction, MB-MB/FE-FE/MB-FE contacts have been established based on contacting surfaces [22]. This defines the human body's contact with the seat cushion, seatback, and floor. Contact interactions are defined between a master surface and a slave surface. Select groups of multibody (MB) surfaces are used as master (planes and ellipsoids) and slave (ellipsoids) in each contact, such as feet contact with floor, pelvis with seat pan and torso with backrest. All contacting surfaces may penetrate one another in this model. The penetration determines the equivalent elastic contact force. For all contacts, linear stiffness and damping coefficients are defined capturing the compression of human tissues and the seat. Thus the contact model captures compliance in compression taking into account the 3D geometry of body and seat. Transmission of shear forces was initially modelled using stick-slip friction [16]. However this proved to be imprecise in reproducing vibration transmission. Hence we removed the friction from the seat contact, and replaced this by point restraints acting orthogonal to the contact surfaces. These point restraints capture seat, muscle, fat and skin shear deformation. These are currently defined as linear force-deflection characteristics with stiffness in N/m and damping in Ns/m and a limited force, to allow contact slip.

*Joint stabilization:* Joint compliance models are used to predict the deformation of bony segments across a variety of body joints [23]. In advanced biomechanical models, the intervertebral joints, are generally modelled as 6 DoF joints allowing compression, shear, and rotation. The formulation of full 6 DoF of each body joint would require the tuning of many of parameters, and would increase the computational demand. By reducing the joint DoF, the model complexity can be reduced. In EHM, joints are efficiently modelled to capture human movements and are kept to be minimum possible. Therefore, most bodies are interconnected by spherical or revolute joints rather than linear (translational) springs and dampers. To re-create the joints' muscles, restraint cardan feature is incorporated. Three torsional parallel springs and dampers that link two bodies make up the Cardan restraint. The torques are influenced by the Cardan angles, which express how the relevant restraint coordinate systems are oriented in relation to one another. Such rotational spring-damper models performed quite well in fitting the data but did not capture the static posture maintenance well. Hence PID feedback joint torque controllers were implemented to maintain the desired body posture. Additionally, Fig. 2 presents the head-in-space control technique, which is used to maintain equilibrium and coordinate head motions during dynamic tasks.

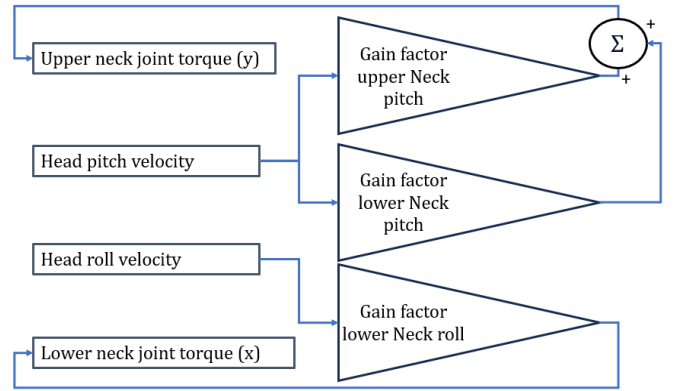

Fig. 2 Head in space (HIS) controller

### III. NUMERICAL SIMULATIONS

The same input signals that the subjects encountered during the experiments were used to replicate the simulations. The seat encountered disturbances at a rate of 0.1941 m/s$^2$ root mean square (RMS) in the x, y, and z directions. According to Fig. 5, between 0-5s, no excitations are provided to the seat, which allows the body to get settled over the seat and reach a static equilibrium position. The vibrational excitations to the seat are given thereafter. This extra 5 sec of simulation time can be avoided using settling method and RESTART technique [24]. For example, a joint position, muscle and actuator controller activation level / restart file can be created at the end of no input equilibrium state and loaded at the beginning of the analysis during the optimization process.

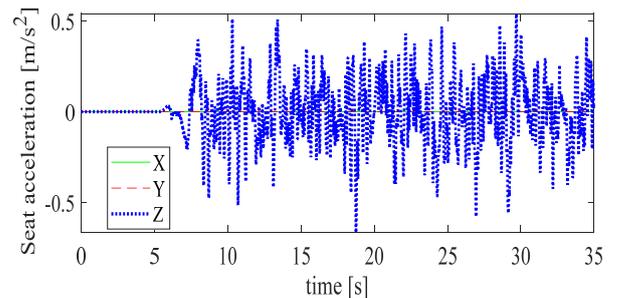

Fig. 3. Seat input excitations.

Both the AHM and the EHM postural stabilization parameters were tuned to this dataset [12]. For EHM, the human body parameters such as mass and inertia value are predefined. It is necessary to determine the unknown parameters, such as hip extension, control gain parameters, stiffness, and damping. One DoF for each revolute joint and translational joint, along with three DoF for the spherical joint. The stiffness and damping represent passive tissue resistance and postural stabilization using muscle feedback and co-contraction. To acquire the model parameters, the iterative parameter optimization technique is used.

### A. Objective criteria

The EHM must respond to the experimental data precisely. To effectively reflect the experimental response, the model must adequately depict the head, pelvis, trunk, and knee in the vertical, fore-aft, and lateral directions. This paper focusses on vibration comfort, and analyzes the transmission of vibrations from seat to pelvis, trunk, head and knees in the frequency domain. For model fitting, we use gains as function of frequency for each body segment and relevant motion direction:

$$\text{Gain} = \frac{\mathcal{F}(s_o)}{\mathcal{F}(s_i)} \quad (1)$$

In this scenario, $s_o$ refers to a human response to a certain body segment in time, such as the vertical displacement of the pelvis or the pitch of the head. $s_i$ stands for the input vibration applied at the seat in time domain. The term $\mathcal{F}$ refers for the Fourier transform, which denotes that the gain is a frequency-domain function. In order for the EHM to be accurate for human reaction, the pertinent gains should have the lowest errors in relation to experimental data for various seat motions. As a result, the criteria, or cost function, for parameter identification are these errors of specific gains. A 0–12 Hz Butterworth band pass filter is implemented for both experimental and model responses. MATLAB's "Tfestimate" function is used to calculate the transfer function estimate. The AHM includes posture controllers to stabilize the body. Some of the free optimization parameters for AHM are activation coefficient (neck, spine, hip, knee and shoulder), extension coefficients (knee, hip) and neck co-contraction. For model validation, the head, pelvis, and trunk in vertical (z, pitch), fore-aft (x, pitch), and lateral (y, roll and yaw) directions must represent the experimental response accurately. Fig. 4 shows a flow chart for the co-simulation and optimization of MATLAB, SIMULINK, and MADYMO. A similar procedure was used for the MADYMO active human model (AHM). The PID integral controller gain settings are selected so that it takes about 3 seconds for joints to reach the desired set point. While the AHM employed a solid FE model for the seat back foam, the EHM used ellipsoid-ellipsoid contacts for the seat back. The time step size is 1E-3 s for EHM whereas due to the presence of finite element (FE), a smaller time step of 5E-5 sec was adopted in the case of AHM. As a result, EHM outperforms AHM by 360 times.

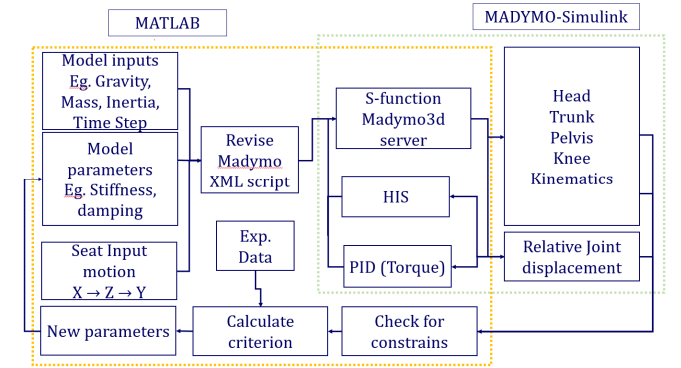

Fig. 4 Co-simulation-optimization flowchart.

### IV. RESULTS

This study analyzes the AHM, the EHM with optimal parameters, and the EHM without integrator and high joint stiffness. Fig. 5 demonstrates that the initially determined posture cannot be maintained in the absence of an integral controller since torso loses touch with the backrest.. In the absence of the integral controller, a head pitch rotation of 6.4 degrees was observed, and also the trunk moved forward. This is due to the low stiffness in neck, trunk and hip joints which was found to best fit the experimental data with the EHM. In the selected erect posture the head and trunk center of gravity are located in front of the joints which stabilize trunk and neck, resulting in forward drift. This was resolved with the integrator controller resulting in a good fit in the frequency domain (see lines EHM in Fig. 6 - Fig. 8). The drift was also resolved with high joint stiffness values, but this resulted in a poor fit in the frequency domain (see lines EHM high stiffness in Fig. 6 - Fig. 8). During the simulations, the existence of a feedback controller will aid in accomplishing any desirable or desired changes in posture. Currently, joint torques are controlled separately to achieve an erect S shape posture, but in the future, this could be integrated via full-body neuromuscular control including vision and vestibular reflexes. The experiments and model results are presented in Fig. 6 - Fig. 8. The accuracy in capturing head and trunk motion was prioritized over pelvic movements during the optimization of the model's parameters due to the complexity of the pelvis and its interactions with other body parts. After parameter optimization, the experimental gain is correctly captured by the AHM and EHM. Additionally, both the AHM and EHM models demonstrate a greater accuracy in capturing the gain of head and trunk motions when compared to the pelvis movements in the experimental data. In several experimental data sets, such as trunk translational gain in lateral direction, EHM outperforms the AHM. Indeed, high stiffness values can be utilized to maintain a prescribed posture during the simulations. By increasing the stiffness, the system becomes less prone to deviating from the desired position. However, there are trade-offs associated with high stiffness values. One of the main drawbacks is increased joint stiffness.

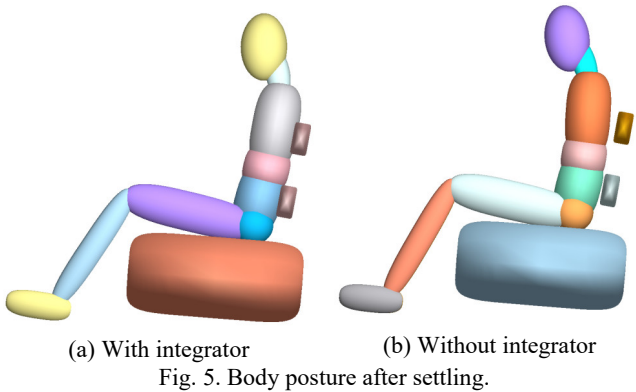

(a) With integrator      (b) Without integrator
Fig. 5. Body posture after settling.

When the stiffness is set too high, it can restrict the natural movement of the joints, leading to decreased flexibility and range of motion.

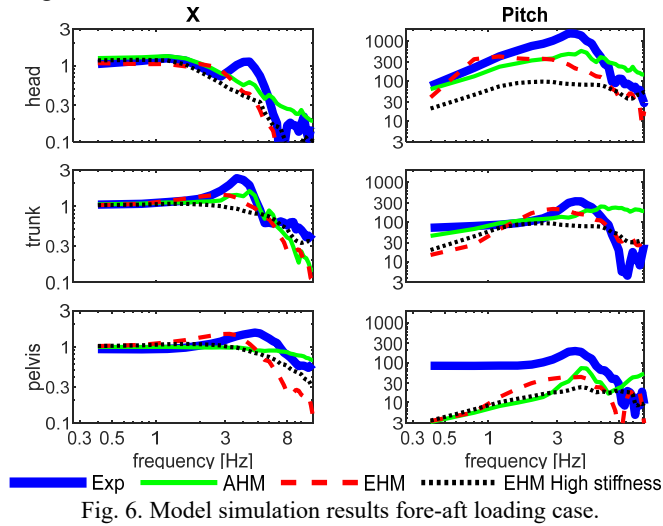

Fig. 6. Model simulation results fore-aft loading case.

Additionally, high stiffness values resulted in decreased gains, particularly in rotational responsiveness., as seen in Fig. 6 - Fig. 8 [Units: translational gain-(m/m) and rotational gain (deg/m)]. As a result, determining the optimum stiffness, damping, and integral values becomes critical in model validation. Adjusting the stiffness within an optimal range can help maintain the prescribed posture while still allowing for efficient movement and control.

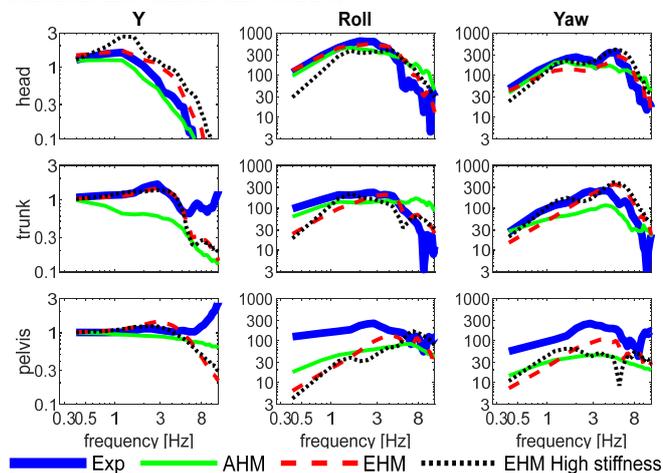

Fig. 7. Model simulation results lateral loading case.

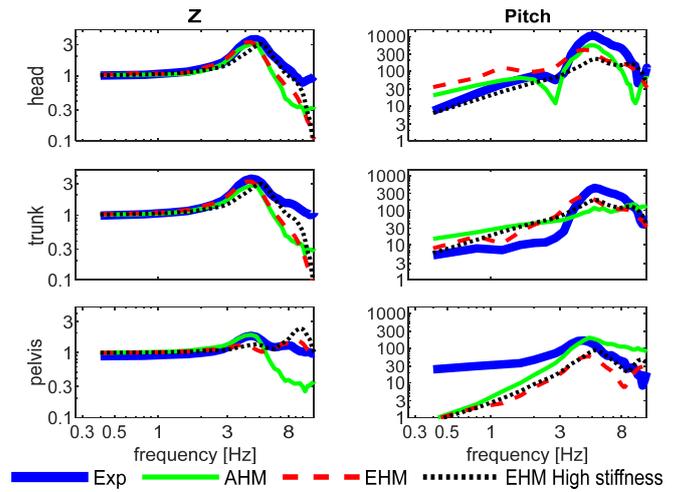

Fig. 8. Model simulation results vertical loading case.

Active muscle controllers will be used in the future to record reflexes, posture corrections, and complicated feedback models, such as proprioceptive, vestibular, and visual motion perception. This will make it possible to create cutting-edge control algorithms for autonomous vehicles utilizing the current EHM, thereby enhancing user comfort.

## V. CONCLUSION

The process of active joint torque for body stabilization is a complex one that involves multiple muscles and joints working together in a coordinated manner. This study emphasizes the significance of head in space control and active feedback for preserving postural stability when seated in a moving vehicle. While advanced 3D non-linear full body models are effective for impact and comfort simulation, they are computationally demanding (360 times EHM), making them less practical for studying long time periods in driving comfort. Therefore, this paper evaluates the effectiveness of simplified 3D full body human models to capture vibration comfort and proposes an efficient seated human body model that can be used to improve the real time comfort and safety of occupants. The study also explores how linear spring-damper models with torque controllers can approximate reflexive postural stabilization. Future research will take into account the sensory input model, which instructs the brain to send signals to the muscles instructing them to contract or relax. Designers and researchers in the automotive and seating industries can benefit from the valuable insights provided by the current study's results, which have significant implications for improving the comfort of occupants.

ACKNOWLEDGMENT

We acknowledge the support of Toyota Motor Corporation.